# A PROTOTYPE OF UPGRADING BEPC CONTROL SYSTEM

C.H. WANG, X. C. KONG, G. LEI, S.F. XU, Q. LE, and J. ZHAO
IHEP, BEIJING, 100039, P.R. CHINA

Abstract

We will use EPICS toolkit [1] to build a prototype for upgrading BEPC control system. The purposes are threefold: (1) Setup a network based distributed control system with EPICS. (2) Study some front-end control technology. (3) Merge the existing control systems into EPICS. The hardware architecture consists of three physical layers. The front-end layer consists of VME crate, PowerPC, VME I/O modules and interface boards (VME-CAMAC and Fieldbus). The back-end layer is a SUN workstation, which is used for application development. The two layers are connected via the middle layer which is 100M Ethernet. In order to preserve our investment in the hardware, the existing CAMAC hardware will remain. The CAMAC will be connected to the VME via CAMAC serial highway [2]. The operator interface will be developed with EPICS DM2K. High-level accelerator control applications will be studied as well with EPICS channel access and archive access.

## 1 INTRODUCTION

BEPC consists of a 200 m long linac and storage ring with a circumference of 240.4 m. It can produce some extremely important high energy physics and synchrotron radiation.

The storage ring control system was transplanted from SPEAR in 1987. The upgrade of this system in 1994 was to replace a couple of very old systems, which were either already obsolete or unsupported (VAX750, Grinnell, VCC). The upgraded system has worked reliably for us.

But this system after the upgrade still has the following shortcomings: (1). it is heavily dependent on the VMS operating system and CAMAC, (2). it is not open to the option of non- CAMAC hardware such as VME, (3). There is a communication bottleneck because real-time data reside in shared memory on a single computer, and (4). the operator interface is intimately bound to the application programs and is primitive by today's standards.

Because the commissioning of BEPCII will be different from BEPC, we are unable to use the FORTH-based OPI and the existing real-time database. Therefore, we will modify and adopt the control system used at other labs for use with BEPCII. Except some subsystems, the control system hardware and software architecture will be upgraded.

The linac control is different from the ring. It uses PC with remote I/O via RS422 to control the devices. The console in the ring only can access some dada such as the Phase in the linac via the network.

## 2 REQUIRMENTS AND OPTIONS

### 2.1 Requirements

Our motivation for doing the upgrade of BEPC is to meet BEPCII [3] requirements. BEPCII will construct double rings in the current tunnel. BEPCII will add some high voltage switch power supplies in each ring. There are about 400 magnet power supplies distributed around the rings and transport lines of BEPCII. So, we need to build new power supply control system on the rings. In order to preserve our investment, the CAMAC in the transport line will remain. We will take a power supply as an example in the prototype and implement the control of the power supply with EPICS. For the requirements of the power supply control are:

- Current setting synchronously without beam loss.
- the fastest time period is 30ms/per step setting.

### 2.2 Options

Several proposals were discussed on how to meet requirements. The proposals submitted consisted of:

- VME + Fieldbus + Intelligent Device controller.
- VME + direct I/O.

We did some market survey for the fieldbus such as Canbus, DeviceNet, ControlNet and Profibus. We think that Canbus only has two-layer protocol supported with 1Mbps speed; it's hard to develop software. SSRF (Shanghai) already developed the DeviceNet communication software between IOCs and device controller. Although ControlNet is faster than DeviceNet, they both have a problem with signal synchrotron processing. So, it is suitable for slow control such as vacuum system. The power supplies on the rings can be directly controlled by VME I/O modules.

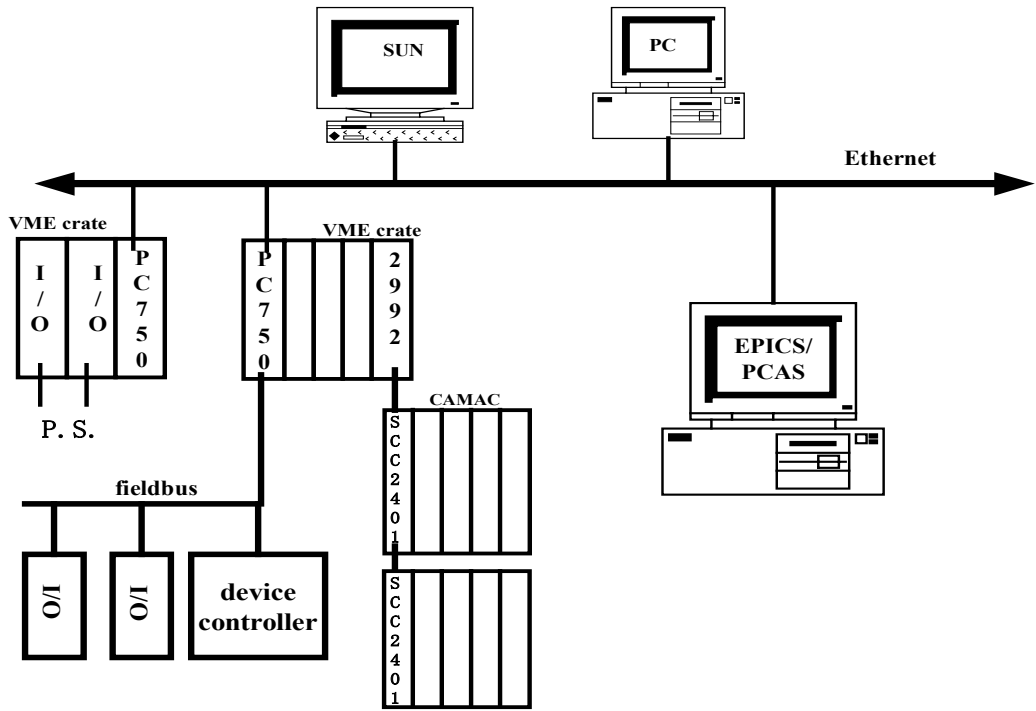

Figure 1: the hardware architecture of the prototype

We are also considering ControlNet for the vacuum control because EPICS already supported ControlNet. We can get the ControlNet driver from the vacuum control group of SNS/BNL.

## 3 ARCHITECTURE OF THE PROTOTYPE

The prototype consists of a Sun workstation, a PC and VMEbus systems as shown in figure 1. The Sun workstation is used as development. The PC is used as OPI. There are three ways to control the devices. One is VME-fieldbus interface controller and a device controller. Another is VME direct I/O module. Another more is VME-CAMAC interface module. In addition, we will install EPICS/PCAS on the PCs running LabView for merge the PC systems into EPICS.

Our main goals for the prototype are: (1)Setup hardware and software architecture with EPICS, (2)Study key technology of the control system integration, (3) Merge the existing PCs and CAMAC into EPICS.

## 4 PLANS

We plan to construct EPICS Real-time database with CapFast and ASCII files according to the BEPCII database name convention. We will build new operator screens with DM2k. We will do EPICS applications on the prototype control system for the power supplies and measure the precision and update rate of acquisition signals. We also will develop the application for the vacuum with VME-fieldbus. Meanwhile we will develop high-level accelerator control applications.

Finally, we will test new IOCs on some subsystems. The power supply system and vacuum system will be first. Then, other subsystems can also be implemented with EPICS soon.

## 5 ACKNOWLEDGEMENTS

We would like to thank all teachers of EPICS seminar and EPICS workshop in Beijing in September for their constructive suggestions and helpful discussions.

## REFERENCES

[1] http://www.aps.anl.gov/epics/index.php.
[2] Karen S. Nolker, Hamid Shaoee, Willaim A. Watson III, Marty Wise, "The CEBAF Accelerator Control System: Migrating from a TACL to an EPICS Based System".
[3] "Feasibility Study Report on BEPCII", Beijing, April, 2001.